\documentclass[slac_one]{revtex4}
\usepackage{graphicx}
\usepackage{fancyhdr}
\begin{document}
\title{{\small{2005 ALCPG \& ILC Workshops - Snowmass, U.S.A.}}\\ %% Please keep this conference title here
\vspace{12pt}
Towards a precise measurement of the top quark Yukawa coupling at the ILC} %% Paper title goes here
\author{A. Juste}
\affiliation{FNAL, Batavia, IL 60510, USA}

\begin{abstract}
A precise measurement of the top quark Yukawa coupling is of great importance, since it may shed light on
the mechanism of EWSB. We study the prospects of such measurement 
during the first phase of the ILC at $\sqrt{s}=500$ GeV, focusing in particular on recent 
theoretical developments as well as the potential benefits of beam polarization. It is shown that both
yield improvements that could possibly lead to a measurement competitive with the LHC.
\end{abstract}
\maketitle
\thispagestyle{fancy}
\section{INTRODUCTION} % Section title should be in all capitals.
The top quark Yukawa coupling ($\lambda_t$) is the largest coupling of the Higgs boson to
fermions. A precise measurement of it is very important since the it may help unravel the secrets of 
the Electroweak Symmetry Breaking (EWSB) mechanism, in which the top quark could possibly
play a key role. For $m_h<2m_t$, a direct measurement of $\lambda_t$ is 
possible via associated $t\bar{t}h$ production, both at the LHC and a future $e^+e^-$ International 
Linear Collider (ILC). At the LHC, the expected accuracy~\cite{lhclc} is $\delta\lambda_t/\lambda_t\sim 12-15\%$ for 
$m_h\sim 120-200$ GeV, assuming an integrated luminosity of 300 fb$^{-1}$. Existing feasibility studies at the ILC~\cite{juste1} 
predict an accuracy of $\delta\lambda_t/\lambda_t\sim 6-10\%$ for $m_h\sim 120-190$ GeV, assuming $\sqrt{s}=800$ GeV and 1000 fb$^{-1}$.
However, currently the baseline design for the ILC only contemplates a maximum center-of-mass energy of
500 GeV. Therefore, it is very relevant to explore the prospects of this measurement
during the first phase of the ILC, especially in view of the limited accuracy expected at the LHC: for a number of years, 
the combination of results from the LHC and ILC would yield the most precise determination of $\lambda_t$.

%\section{DIRECT MEASUREMENT OF THE TOP QUARK YUKAWA COUPLING AT \mbox{\boldmath$\sqrt{s}=500$} GeV}
A preliminary feasibility study at $\sqrt{s}=500$ GeV was performed in Ref.~\cite{juste2}, which we briefly overview here.
Indeed, the measurement of $\lambda_t$ at $\sqrt{s}=500$ GeV is more challenging than at $\sqrt{s}=800$ GeV.
On the one hand, the reduced phase-space leads to a large reduction in $\sigma_{t\bar{t}h}$
(e.g. $\sigma^{Born}_{t\bar{t}h}\simeq 0.16(2.5)$ fb at $\sqrt{s}=500(800)$ GeV, for $m_h=120$ GeV). On the other hand, 
the cross section for many background processes is significantly increased. This analysis assumed 
$m_h=120$ GeV and focused on the $t\bar{t}h\rightarrow (\ell\nu b)(jjb)(b\bar{b})$ decay channel ($BR\sim 30\%$).
The dominant background is $t\bar{t}jj$, followed by $t\bar{t}Z$, although other non-interfering
backgrounds (e.g. $W^+W^-$) were also considered. Signal and backgrounds were processed through a fast detector simulation.
After basic preselection cuts, the signal efficiency was found to be $\simeq 50\%$ and the $S:B\simeq 1:100$.
In order to increase the sensitivity, a multivariate analysis using a Neural Network (NN) with 23 variables was performed.
The final selection consisted on an optimized cut on the NN distribution. Assuming an integrated luminosity of
1000 fb$^{-1}$, the expected total number of signal and background events was 11 and 51, respectively, resulting
in $(\delta\lambda_t/\lambda_t)_{stat}\simeq 33\%$. Based on previous experience~\cite{juste1}, the addition of the 
fully hadronic decay channel was expected to ultimately lead to $(\delta\lambda_t/\lambda_t)_{stat}\simeq 23\%$. 
While this analysis is already rather sophisticated,
significant improvements are expected from e.g. the usage of a more efficient $b$-tagging algorithm or a more optimal 
treatment of the kinematic information. In the next sections we discuss additional sources of improvement which are 
currently under investigation.

\section{THE IMPACT OF RESUMMATION EFFECTS}

\begin{figure*}[t]
\centering
\includegraphics[width=5.5cm]{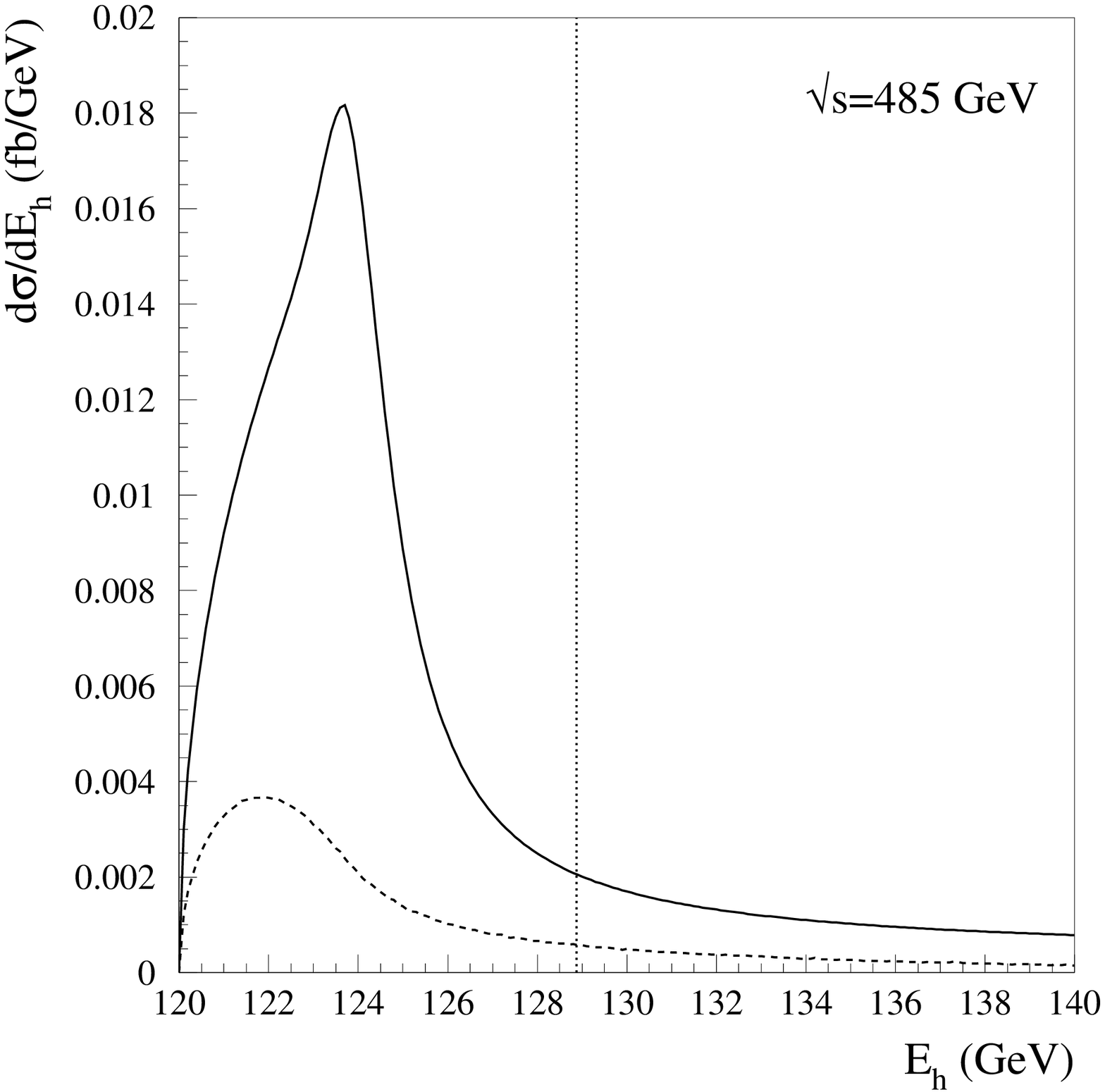}
\includegraphics[width=5.5cm]{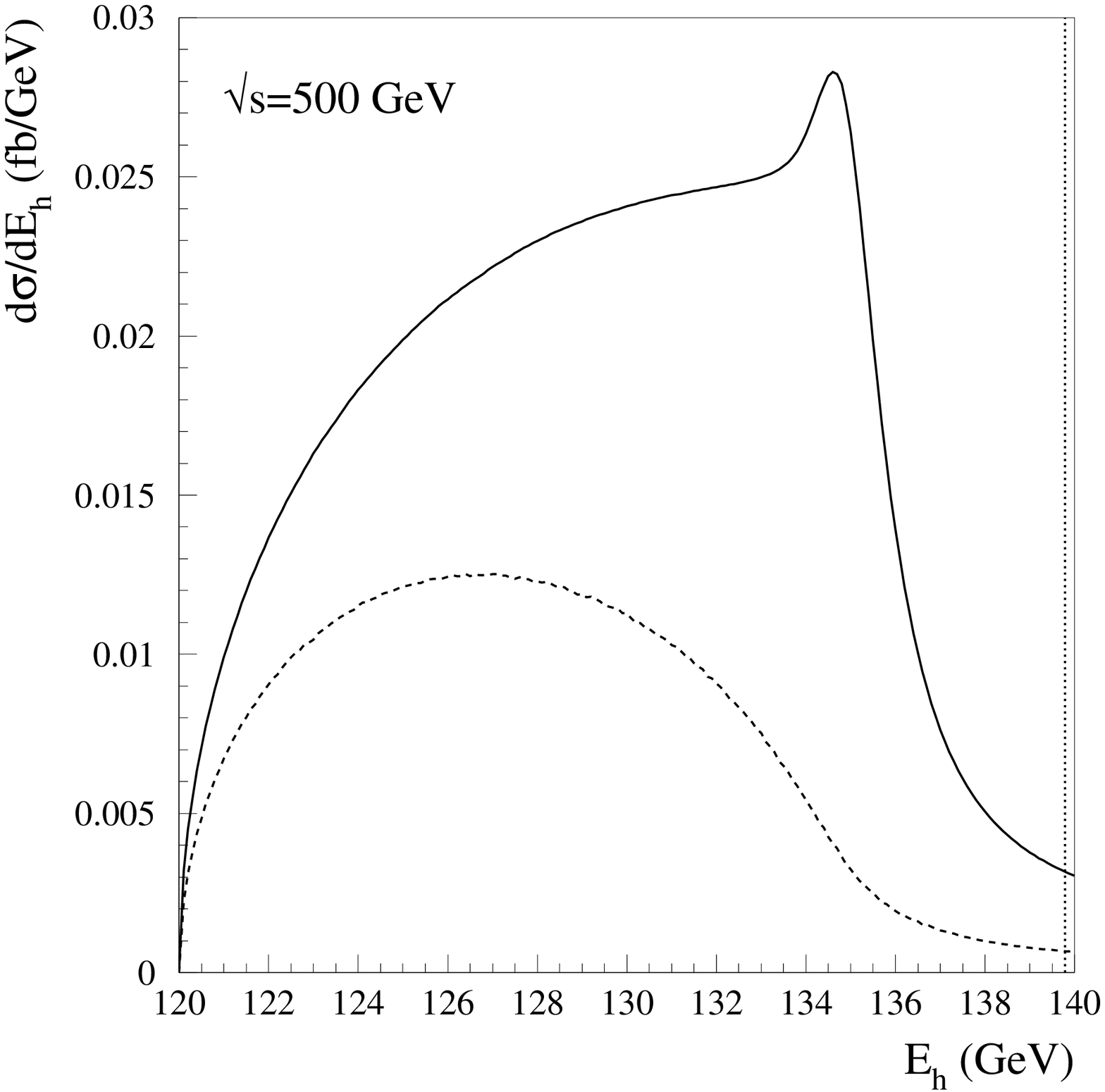}
\includegraphics[width=5.5cm]{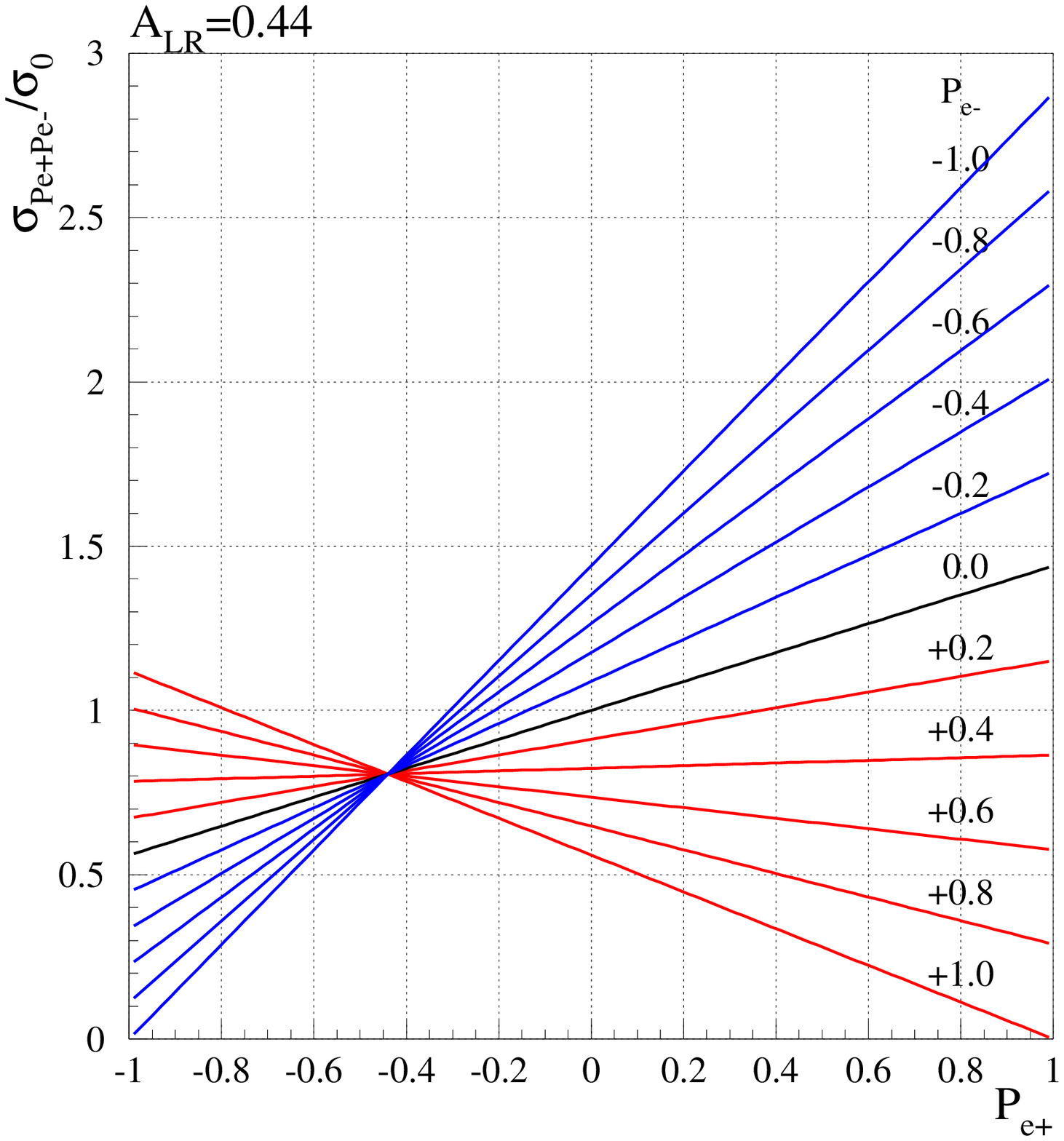}
\caption{Left and center: comparison of the Born (dashed) and NLL (solid) $d\sigma_{t\bar{t}h}/dE_h$ for different values of $\sqrt{s}$, assuming
$m_t^{1S}=180$ GeV and $m_h=120$ GeV. The dotted line indicates the value of $E^{max}_h$.
Right: ratio of polarized to unpolarized cross section for different values of ($P_{e^-},P_{e^+}$).}
\label{combo}
\end{figure*}

The precise measurement of $\lambda_t$ requires accurate theoretical predictions for $\sigma_{t\bar{t}h}$. Currently, one-loop
QCD and electroweak corrections are available. However, at $\sqrt{s}=500$ GeV and for $m_h\geq 120$ GeV,
the kinematic region where the $t\bar{t}$ system is non-relativistic dominates. As discussed in Ref.~\cite{andre}, in this regime 
Coulomb singularities are important and need to be resummed within the framework of the vNRQCD effective theory, leading to
large enhancements factors in the cross section relative to the Born level. At the ILC, because of ISR and beamstrahlung (BS),
the event-by-event center-of-mass energy ($\sqrt{\hat{s}}$) will be lower than the nominal one, 
thus bringing the $t\bar{t}$ system deeper into the non-relativistic regime. In order to compute the expected $\sigma_{t\bar{t}h}$ including these effects,
the 11-fold Born differential cross section for $e^+e^- \rightarrow t\bar{t}h \rightarrow W^+bW^-\bar{b}h$ was multiplied 
by a K-factor defined as $K(E_h,\sqrt{\hat{s}})=(d\sigma_{t\bar{t}h}^{NLL}/dE_h)/(d\sigma_{t\bar{t}h}^{Born}/dE_h)$,
where $E_h$ stands for the Higgs boson energy in the $e^+e^-$ rest-frame, and then folded with ISR and BS structure functions.
The NLL differential cross section was kindly provided by the authors of Ref.~\cite{andre}.
Fig.~\ref{combo}(left and center) compares the Born (for off-shell top quarks) and NLL differential cross sections 
for different values of $\sqrt{\hat{s}}$, assuming $m_t^{1S}=180$ GeV and 
$m_h=120$ GeV. The ratio of these two curves defines $K(E_h,\sqrt{\hat{s}})$ and can be significantly
larger than 1, especially for low values of $\sqrt{\hat{s}}$. Since the NLL prediction is only valid for 
$E_h\leq E^{max}_h$ (where $E^{max}_h$  effectively corresponds to a cut on the top quark velocity in the $t\bar{t}$ rest-frame of $\beta_t<0.2$), 
we currently set $K(E_h,\sqrt{\hat{s}})=1$ for $E_h>E^{max}_h$, although in practice, it should be possible to use
$K(E_h,\sqrt{\hat{s}})=(d\sigma_{t\bar{t}h}^{O(\alpha_s)}/dE_h)/(d\sigma_{t\bar{t}h}^{Born}/dE_h)$.
Table~\ref{sigmatth} compares the predicted Born and ``NLL-improved'' $\sigma_{t\bar{t}h}$
for different scenarios, illustrating the large impact of radiative effects in the initial state.
This underscores the importance of being able to predict these effects to the percent level. 
While the impact of ISR cannot be reduced, it might be possible to find an optimal operating point of the accelerator, as far as this measurement
is concerned, in terms of BS and total integrated luminosity. 
Finally, it is found that, for $m_h=120$ GeV, resummation effects can increase $\sigma_{t\bar{t}h}$ by a
factor of $\sim 2.4$ with respect to the Born cross section, used in the previous feasibility study.

\begin{table}[t]
\begin{center}
\caption{Comparison of the Born and NLL $\sigma_{t\bar{t}h}$ for different scenarios regarding radiative effects in the initial state.}
\begin{tabular}{|c|c|c|c|}
\hline \textbf{(ISR,BS)}  & \textbf{\boldmath$\sigma_{t\bar{t}h}$ (fb) (Born)} & \textbf{\boldmath$\sigma_{t\bar{t}h}$ (fb) (``NLL-improved'')} & \textbf{Enhancement factor} \\
\hline 
(off,off) & 0.157(1) & 0.357(2) & 2.27 \\
\hline 
(off,on) & 0.106(1) & 0.252(3) & 2.38 \\
\hline 
(on,on) & 0.0735(8) & 0.179(2) & 2.44 \\
\hline
\end{tabular}
\label{sigmatth}
\end{center}
\end{table}

\section{THE IMPACT OF BEAM POLARIZATION}

So far, all feasibility studies of this measurement have assumed unpolarized beams. Currently, the baseline design for the ILC only includes
longitudinal electron beam polarization ($|P_{e^-}|\simeq 0.8$). Positron beam polarization ($|P_{e^+}|\simeq 0.6$) is considered as an option.
The ratio of the polarized cross section (for arbitrary longitudinal beam polarization) 
($\sigma_{P_{e^-}P_{e^+}}$) to the unpolarized cross section ($\sigma_0$) is given by
$\sigma_{P_{e^-}P_{e^+}}/\sigma_0 = (1-P_{e^-}P_{e^+})(1-P_{eff}A_{LR})$, where $P_{eff}=(P_{e^-}-P_{e^+})/(1-P_{e^-}P_{e^+})$ denotes the
``effective polarization'' and $A_{LR}$ is the ``left-right asymmetry'' of the process of interest~\cite{gudi}.
Therefore, two potential enhancement factors can in principle be exploited: the first one requires having both beams polarized, the second
one requires $A_{LR}\neq 0$ and a judicious choice of the signs of $P_{e^-}$ and $P_{e^+}$ in order to have $P_{eff}A_{LR}<0$.
In the case of SM $t\bar{t}h$ production, $A_{LR} \simeq 0.44$, essentially independent of $\sqrt{s}$ 
in the range $\sim 0.5-1.0$ TeV. Assuming $(A_{LR})_{SM}$, Fig.~\ref{combo}(right) shows the cross section enhancement factor as a function 
of $P_{e^+}$, for different values of $P_{e^-}$. The optimal (realistic) operating point would be $(P_{e^-},P_{e^+})=(-0.8,+0.6)$, achieving an
increase in $\sigma_{t\bar{t}h}$ by a factor of $\simeq 2.1$ with respect to the unpolarized case. Unfortunately, this choice does not
help reduce the dominant background, which is increased by a similar factor. Nevertheless, the net result is still an improvement
in the statistical precision on $\lambda_t$ by $\sim 45\%$, which would be an argument in favor of including positron polarization 
in the baseline design. For $(P_{e^-},P_{e^+})=(-0.8,0)$, only a modest increase in $\sigma_{t\bar{t}h}$ by a factor of $\sim 1.3$ would be achieved.
It is important to realize that, in order to choose the sign of $P_{e^-}$ and $P_{e^+}$, it is necessary to know the sign of $A_{LR}$.
Anomalous couplings in the $tt\gamma$ and $ttZ$ vertices could possibly lead to
deviations in $A_{LR}$ from the SM prediction. Unfortunately, due to the limited precision
in the measurement of the $ttZ$ couplings~\cite{baur}, the LHC is not expected to provide any useful constraints on the sign of $A_{LR}$.
Therefore, at the ILC the first step should be to perform measurements of the polarized $\sigma_{t\bar{t}}$ in order to determine
the sign of $A_{LR}$, and thus fix the signs of $P_{e^-}$ and $P_{e^+}$ (the magnitudes should be largest possible). On the other hand, admittedly the 
measurement of $\lambda_t$ requires a percent-level and model-independent determination of the $t\bar{t}\gamma$ and $t\bar{t}Z$ couplings, which
typically benefits from changing the beam polarization. Therefore, it would be desirable to optimize the running strategy to maintain the largest
possible $\sigma_{t\bar{t}h}$, needed for a precise measurement of $\lambda_t$, while meeting the precision goals for measurements of top quark couplings.

\section{CONCLUSIONS}
We have studied the prospects of a precise measurement of the top quark Yukawa coupling during the first phase of the ILC.
Taking into consideration an existing feasibility study, and the additional enhancement factors to $\sigma_{t\bar{t}h}$ discussed 
here, we anticipate a precision of $(\delta\lambda_t/\lambda_t)_{stat}\sim 10\%$ for $m_h=120$ GeV, assuming $\sqrt{s}=500$ GeV and 1000 fb$^{-1}$.

\begin{acknowledgments}
The author would like to thank Andr\'{e} Hoang for useful discussions and for making available the NLL predictions used in this study.
\end{acknowledgments}

%\begin{thebibliography}{9}   % Use for  1-9  references

\end{document}